\documentclass[%
 reprint,
superscriptaddress,
amsmath,amssymb,
aps,
]{revtex4-1}

\usepackage{graphicx}% Include Fig. files
\usepackage{dcolumn}% Align table columns on decimal point
\usepackage{bm}% bold math
\newcommand{\upsub}[1]{\sb{\mathrm{#1}}}
\newcommand{\upsup}[1]{\sp{\mathrm{#1}}}

\begingroup\lccode`~=`_\lowercase{\endgroup\let~\upsub}
\begingroup\lccode`~=`^\lowercase{\endgroup\let~\upsup}

\AtBeginDocument{%
  \catcode`_=12 \catcode`^=12
  \mathcode`_="8000
  \mathcode`^="8000
}

\begin{document}
\preprint{APS/123-QED}

%\title{Second-harmonic generation in a super-crystal with giant refraction}

\title{Constraint-free wavelength conversion supported by giant optical refraction \\ in a 3D  perovskite supercrystal}

\author{Ludovica Falsi}
\affiliation{Dipartimento di Fisica, Universit\`{a} di Roma ``La Sapienza'', 00185 Rome, Italy}
\affiliation{Dipartimento S.B.A.I., Sezione di Fisica, ``Sapienza'' Universit\`{a} di Roma, I-00161 Roma, Italy}
\author{Luca Tartara}
\affiliation{Dipartimento di Ingegneria Industriale e dell'Informazione, Universit\`{a} di Pavia, I-27100 Pavia, Italy}
\author{Fabrizio Di Mei}
\affiliation{Dipartimento di Fisica, Universit\`{a} di Roma ``La Sapienza'', 00185 Rome, Italy}
\author{Mariano Flammini}
\affiliation{Dipartimento di Fisica, Universit\`{a} di Roma ``La Sapienza'', 00185 Rome, Italy}
\author{Jacopo Parravicini}
\affiliation{Dipartimento di Scienza dei Materiali, Universit\`{a} di Milano-Bicocca, I-20125 Milano, Italy}%
%\affiliation{Universit\`{a} di Milano Bicocca}
%%
\author{Davide Pierangeli}
\affiliation{Dipartimento di Fisica, Universit\`{a} di Roma ``La Sapienza'', 00185 Rome, Italy}
%%%
%%
\author{Gianbattista Parravicini}
\affiliation{Dipartimento di Fisica, Universit\`{a} di Pavia, I-27100 Pavia, Italy}%
%\affiliation{Universit\`{a} di Pavia}
%%%
\author{Feifei Xin}
\affiliation{Dipartimento di Fisica, Universit\`{a} di Roma ``La Sapienza'', 00185 Rome, Italy}
\affiliation{College of Physics and Materials Science, Tianjin Normal University, Tianjin, China, 300387}

\author{Paolo Di Porto}
\affiliation{Dipartimento di Fisica, Universit\`{a} di Roma ``La Sapienza'', 00185 Rome, Italy}

\author{Aharon J. Agranat}
\affiliation{Applied Physics Department, Hebrew University of Jerusalem, IL-91904 Jerusalem, Israel}
%%%
\author{Eugenio DelRe}
\affiliation{Dipartimento di Fisica, Universit\`{a} di Roma ``La Sapienza'', 00185 Rome, Italy}
\affiliation{ISC-CNR, Universit\`a di Roma ``La Sapienza'', 00185 Rome, Italy}

\date{\today}

\begin{abstract}
\noindent     
\textbf{Abstract} Nonlinear response in a material increases with its index of refraction as $n^4$. Commonly, $n \sim$ 1 so that diffraction, dispersion, and chromatic walk-off limit nonlinear scattering. Ferroelectric crystals with a periodic 3D polarization structure overcome some of these constraints through versatile Cherenkov and quasi-phase-matching mechanisms. Three-dimensional self-structuring can also lead to a giant optical refraction. Here, we perform second-harmonic-generation experiments in KTN:Li in conditions of giant broadband refraction. Enhanced response causes wavelength conversion to occur in the form of bulk Cherenkov radiation without diffraction and chromatic walk-off, even in the presence of strong  wave-vector mismatch and highly focused beams. The process occurs  with a wide spectral acceptance of more than 100 nm in the near infrared spectrum, an ultra-wide angular acceptance of up to $\pm 40^{\circ}$, with no polarization selectivity, and can be tuned to allow bulk supercontinuum generation. Results pave the way to highly efficient and adaptable nonlinear optical devices with the promise of single-photon-to-single-photon nonlinear optics.

\end{abstract}
\pacs{Valid PACS appear here}

\maketitle %% required
%%%%%%%%%%%%%%%%%%%%%%%%%%%%%%%%%%%%%%%%%

\section*{Introduction}
\noindent
Frequency conversion and parametric amplification are fundamental ingredients for a wide family of applications, including light sources, detection, optical processing, and quantum-state-generation \cite{Boyd2008,Shen1984, Loudon2010, Brown2003}. For quantum technology, a versatile and super-efficient nonlinear process is the key to photon-based quantum computing \cite{Tiecke2014,Reiserer2014,Chang2014}.  In most schemes, optical nonlinearity can only be effectively harnessed  when the coupling mechanism is driven by a cumulative wave interaction based on constructive interference.  This imposes specific constraints on the available conversion schemes, so-called phase-matching conditions that depend both on the polarization, wavelength, and direction of propagation of the interacting waves and on the specific nonlinear susceptibility of the medium \cite{Boyd2008}.\\

 These constraints can be overcome in engineered ferroelectric crystals with a full 3D periodic spontaneous polarization distribution through quasi-phase-matching \cite{Wei2018,Xu2018,Zhang2019,Stoica2019,Liu2019} and Cherenkov phase-matching \cite{Jelley1958,Mathieu1969,Tien1970,Zhang2008,Sheng2010,Sheng2012,Roppo2013,Ni2016}.
3D lattices of spontaneous polarization also occur naturally in nanodisordered ferroelectrics in the form of supercrystals \cite{Pierangeli2016}, in which case the highly ordered domain mosaic also leads to giant broadband optical refraction \cite{DiMei2018}.  This has a direct effect on nonlinear scattering.  Considering material polarization $P$ in terms of  the Taylor series expansion in the propagating optical field $E_{opt}$, i.e., $P=\epsilon_0\left(\chi^{(1)}E_{opt}+\chi^{(2)}E^2_{opt}+...\right)$ \cite{Armstrong1962,Boyd2008}, the first term describes linear response through the first order susceptibility $\chi^{(1)}= n^2-1$, while higher-order terms describe nonlinear effects.  The validity of the expansion implies $\chi^{(m+1)}/\chi^{(m)} \sim 1/E_{at}$, where $E_{at}$ is the scale of the atomic electric field of the substance. It follows that the intensity of an arbitrary allowed nonlinear scattering processes scales with $(\chi^{(m)})^2(E_{opt})^{2m} \sim (\chi^{(1)}E_{opt})^2(E_{opt}/E_{at})^{2m}$, and the intensity of any higher order scattering processes scales with $n^4$\cite{Loudon2010}. In these term giant refraction, i.e., an index of refraction $n\gg1$ across the visible and near infrared spectrum, forms a direct route to strongly enhanced nonlinear response. \\ 
\noindent

Here we investigate second-harmonic-generation in a ferroelectric supercrystal  manifesting giant refraction.  Enhanced response allows the process to occur through bulk nonlinear Cherenkov radiation even for highly focused non-phase-matched beams, a method to achieve constraint-free wavelength conversion.

\section*{Results and Discussion}
\subsection{Giant Refraction Cherenkov SHG}
\noindent
%\textbf{Giant Refraction Cherenkov SHG.} 
In the paradigm nonlinear optical process, second-harmonic-generation (SHG), waves are generated at wavelength $\lambda/2$ (and angular frequency $2\omega$) by the anharmonic response of dipoles driven by the pump at wavelength $\lambda$ \cite{Kleinman1962}. The process occurs most efficiently when the converted signal interferes constructively with the pump itself, a phase-matching condition that embodies momentum conservation for the interaction.  For any given material, dispersion causes phase-matching to occur naturally in the direction of the pump only for converted light (signal beam) whose wavevector $\mathbf{k}_{2\omega}$ forms a finite angle $\theta_C'$ relative to the pump itself $\mathbf{k}_{\omega}$.  This leads to wavelength-dependent constraints on the process geometry while the wavevector mismatch $\Delta \mathbf{k}=\mathbf{k}_{2\omega}-2\mathbf{k}_{\omega}$  is accompanied by chromatic walk-off \cite{Shen1984}  (see Fig. 1a top panels). Collinear phase-matching ($\Delta \mathbf{k}=0$) can, in turn, be achieved using material birefringence, which introduces wavelength and polarization constraints \cite{Boyd2008}, and quasi-phase-matching, that requires periodic material microstructuring and is also wavelength-selective \cite{Fejer1992,Saltiel2009}.  With $n\gg1$, the angle  at which Cherenkov phase-matching occurs is greatly reduced ($\theta_C'\simeq 0$) so that  chromatic walk-off does not intervene (see Giant refraction Cherenkov phase-matching in Methods and in Fig. 1a (bottom panels)).

The specific geometrical structure of Cherenkov SHG combined with giant refraction is illustrated in Fig. 1b.  The pump  propagates inside the sample along the normal to its input facet irrespective of launch angle $\theta_i$ (left panels) and the Cherenkov SHG copropagates with the pump inside the sample ($\theta_C'\simeq0$, central panels). At the output facet the pump and signal separate at a now finite angle $\theta_C$, as illustrated for the two cases of transverse electric (TE) and trasverse magnetic (TM) polarizations (central and right panels, respectively).  For the TE case, while the pump exits with an angle to the normal $\theta_0=\theta_i$, the Cherenkov SHG forms two beams angled with respect to the pump by $\theta_0\pm \theta_C$, with the same polarization as the pump and on the incidence plane (the xz plane).  In turn, for the TM case, the SHG Cherenkov radiation separates at the output in the orthogonal plane (the yz plane) (see Giant refraction Cherenkov SHG in Methods).

%----Figure1-------%
\begin{figure*}[!ht]
\centering
\includegraphics[width=2.0\columnwidth]{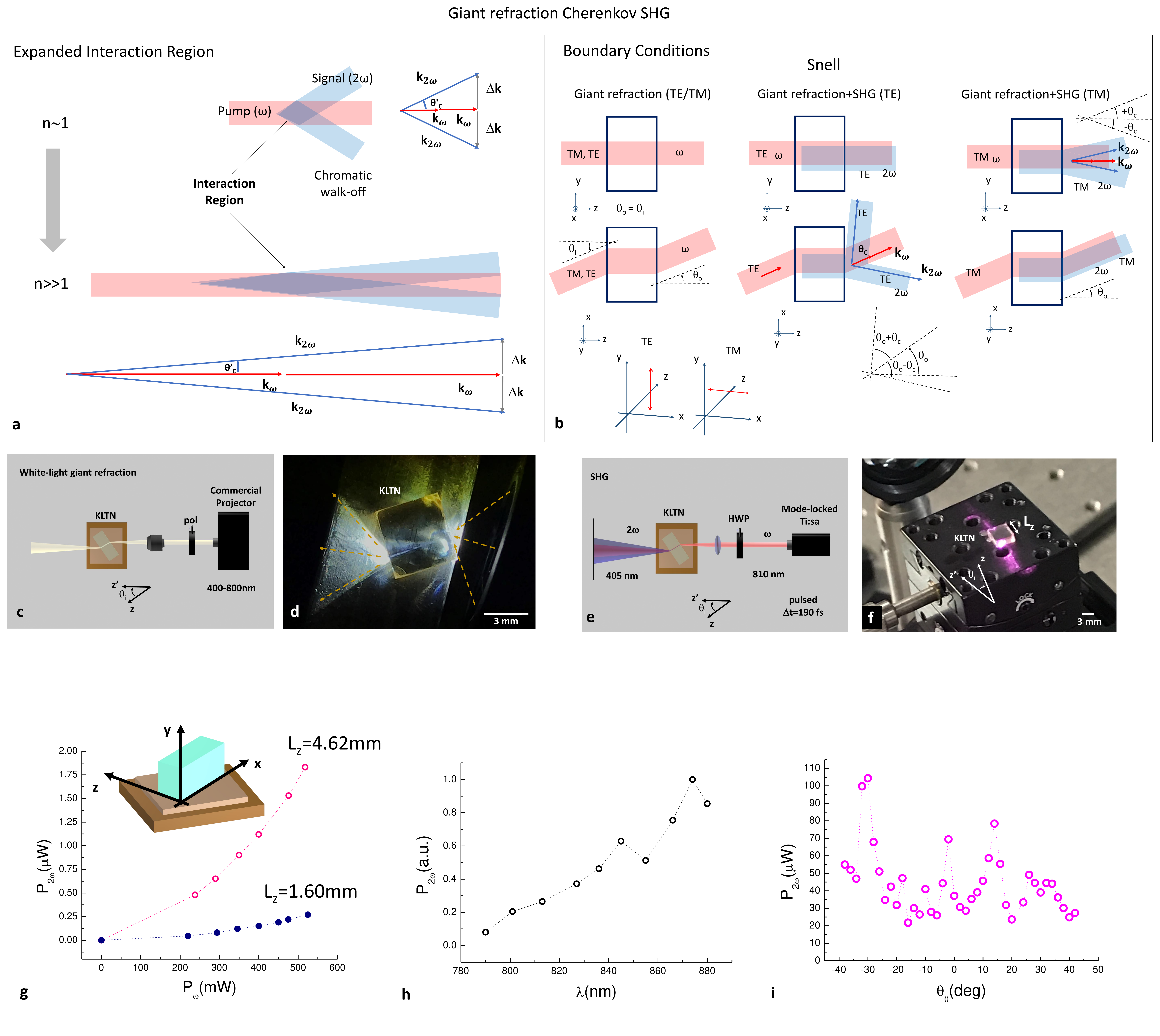}
\caption{\textbf{Giant refraction Cherenkov Second-Harmonic-Generation.} (a) For $n\sim1$, a finite Cherenkov phase-matching $\theta_C'$ leads to a limited beam interaction region associated to a finite beam width and chromatic walk-off. For $n\gg1$, chromatic walk-off $\theta_C' \simeq 0$, so that the interaction region is expanded. (b) Geometry of giant refraction Cherenkov SHG for the TE and TM cases (see Giant refraction Cherenkov SHG Section in Methods). (c) Giant refraction is observed in a nanodisordered KTN:Li crystal cooled 15 K below the $T_C=313$K Curie point using a white-light from a commercial projector, leading to a signature achromatic propagation orthogonal to the input facet (along $z$) irrespective of the launch direction $z'$ (and launch angle $\theta_0$) with diffraction only occuring as the beam leaves the sample (see Giant refraction Experiments Section in Methods). (d) Top view of basic evidence of giant refraction for white light propagation in KTN:Li. (e), (f) SHG is observed using a mode-locked Ti:sa laser (see SHG Setup Section in Methods). (g) Average output SHG power versus pump input power along two different lengths of one sample (sample 1).  Conversion scales with $P_\omega^2$ and with $L_z^2$ as would occur for bulk SHG conversion \cite{Armstrong1962}. Super-broad SHG (h) wavelength and (i) angular acceptance (see Acceptance Section in Methods).      
}
\label{figure1}
\end{figure*}

\begin{figure*}[!ht]
\centering
\includegraphics[width=2\columnwidth]{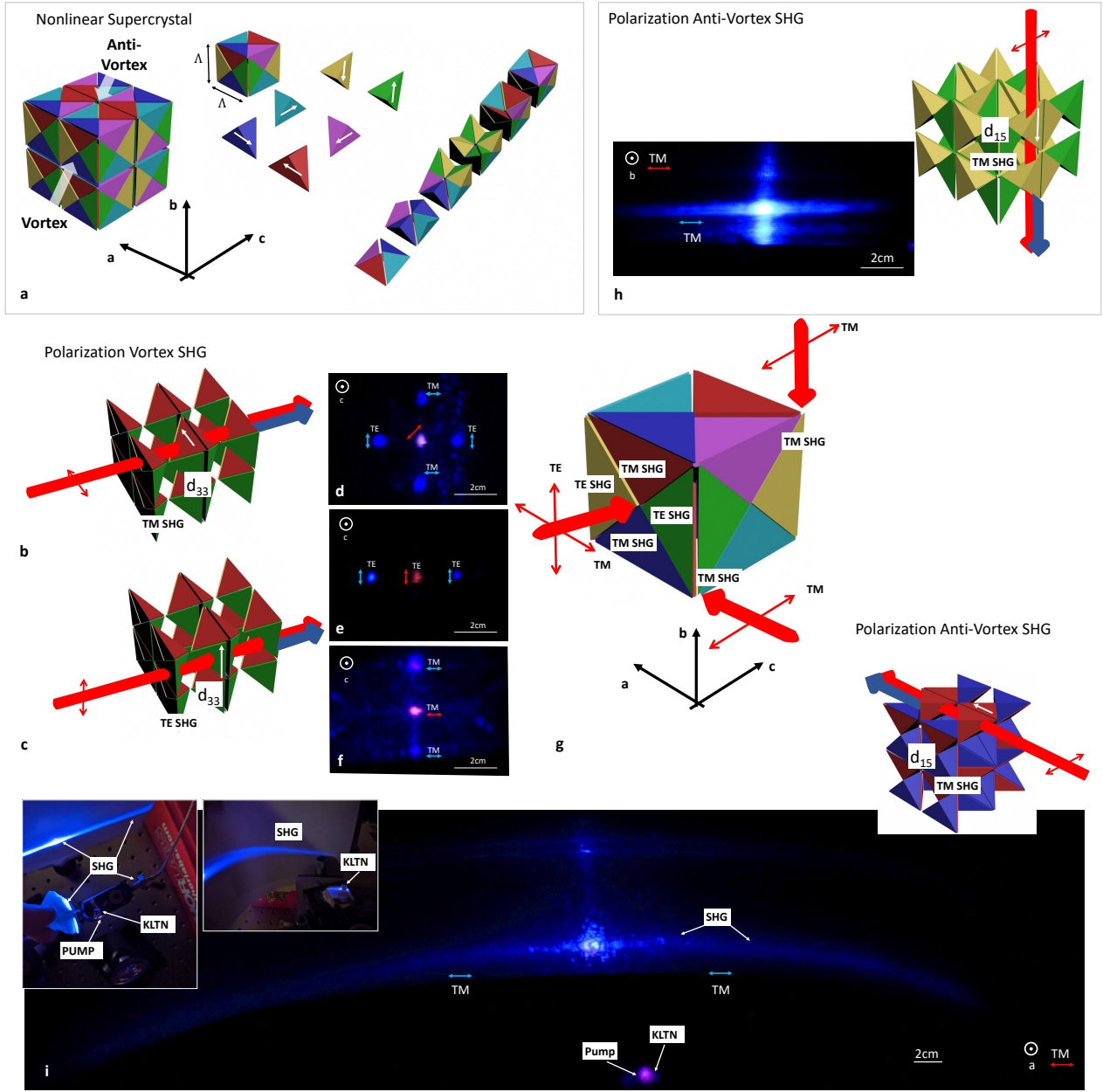}
\caption{\textbf{SHG in a supercrystal.}  (a) Illustration of a specific realization of a supercrystal.  The spontaneous polarization, the white arrows, determines the specific $\chi^{(2)}$ response of each composing tetrahedral (a color coding is implemented).  The actual structure of a fundamental $\Lambda \times \Lambda \times \Lambda$ cube is illustrated through a sequential build-up adding groups of tetrahedral domains. The lattice constant $\Lambda$ ($\simeq 50 \mu m$) is fixed by the built sample growth process (see Materials Section in Methods).  The supercrystal can have many different chiral realizations.  The one illustrated here serves to explain the specific results found, relative to the crystal growth axis $a$.  Indicated are the regions leading to strongest SHG, i.e., the core of polarization vortices in the $ab$ facet and the core of anti-vortices in the $ac$ and $bc$ facets. (b), (c) Non-zero contributions to $\chi^{(2)}$ are illustrated for the TM (b) and TE cases (c). (d)-(f) Output spatial distribution and polarization distribution for a pump polarized at 45 degrees (d), TE (e), and TM (f). (g) Schematic illustration of the experiments for a pump focused into a single polarization-vortex in the $ab$ facet and into an anti-vortex both in the $bc$ and $ac$ facets. (h) SHG for a pump focused in an anti-vortex in the $ac$ facet.  As illustrated in the inset, the only finite contribution to SHG is mediated by $b$ polarized domains through the $d_{15}$ component for TM. (i) SHG for a pump focused in an anti-vortex in the $bc$ facet.  Results are analogous to the $ac$ facet, while here a full angle Cherenkov emission is clearly visible ($2\theta_C=\pi$).  Note how, in distinction to random-phase-matching, no lateral emission is observed \cite{Roppo2010,Ayoub2011,Molina2008}, and all SHG is originating, as expected, solely from the output facet (see inset photographs).}  
\label{figure2}
\end{figure*} 

We perform experiments in two samples of nanodisordered oxide ferroelectric KTN:Li perovskites (see Materials Section in Methods).  These  manifest giant refraction, with record-high  broadband index of refraction ($n>26$) at visible wavelengths.  The effect is associated to the emergence of an  underlying  supercrystal \cite{Pierangeli2016}.  Each lattice site of the supercrystal is the core of a periodic 3D vortex and anti-vortex structure, a mesh of spontaneous polarization that forms below the Curie point (see Supercrystal Preparation Section in Methods).  Typical broadband giant refraction for sample 1 is reported in Figs. 1c, d (see Giant refraction Experiments Section in Methods).  In Figs. 1e, f we illustrate the scheme used to investigate SHG using 190 fs pulses from a mode-locked Ti:Sa source (see SHG Setup Section in Methods).  In Fig. 1g we report output SHG power versus pump input power.  The observed scaling $P_{2\omega} \propto (P_{\omega})^2L_{\mathit z}^2$, where $L_{\mathit z}$ is the length of the sample in the $z$ direction, is reminiscent of the undepleted pump regime of standard SHG \cite{Armstrong1962, Boyd2008}.  The $P_{2\omega}/P_{\omega}$ ratio is independent of input polarization, and output polarization is found to coincide with the input.  In Fig. 1h we report SHG conversion varying the pump wavelength in the available pump spectrum (see Acceptance Section in Methods). As reported in Fig. 1i, SHG conversion is observed for all accessible input angles $\theta_i$ ($=\theta_0$), indicating that the conversion occurs also with no input angular acceptance (see Acceptance Section in Methods).   

\subsection{An underlying 3D nonlinear lattice.}
\noindent
%\textbf{An underlying 3D nonlinear lattice.} 
Wavelength conversion is mediated by the second-order nonlinear susceptibility  response $\chi^{(2)}$ of the KTN:Li perovskite in its noncentrosymmetric tetragonal 4mm state.  In distinction to single-domain  or to quasi-phase-matching schemes, the nonlinear process is mediated by a supercrystal with its specific 3D geometry, giant refraction, and underlying ferroelectric domain structure \cite{Sheng2010,Wang2017}. Hence, while giant refraction causes conversion efficiency to be essentially independent of polarization, input angle, and wavelength, the details of the SHG output strongly depend on input parameters and the supercrystal structure. As illustrated in Fig. 2a, the structure of the 3D supercrystal is a volume lattice of 3D polarization vortices that emerge as the cubic symmetry is broken and polarization charge is screened \cite{Pierangeli2016,DiMei2018}.  The supercrystal forms from the periodic compositional disorder along the growth direction (the $a$ axis).  Each domain has its spontaneous polarization along one of the 6 principal directions (the direction of the spontaneous polarization  is labeled using different colors , see white arrows and colored solids in Fig. 2a). In each domain (of a given color), the corresponding nonlinear susceptibility tensor $d$ depends on its orientation. Consider now the pump focused into a vortex site on the $a,b$ facet of the supercrystal (Fig. 2b, c).  For a TM polarization, most of the component solids lead to a net zero $\chi^{(2)}$ effect, as light experiences a sequence of oppositely polarized tetrahedrals.  The tetrahedrals that dominate $\chi^{(2)}$ response are those with a spontaneous polarization in the $a$ direction, if light propagates along the $c$ direction shifted in the $b$ direction above and  below a single polarization vortex. Here conversion occurs through a sequence of solids with identically oriented polarization  (see Fig. 2b). In the TE case, the situation is analogous, but the SHG signal is now produced for light propagating in the $c$ direction in regions shifted in the $a$ direction in proximity of the vortex (see Fig. 2c).  Focusing the pump on the $ab$ facet into a polarization vortex leads to the output intensity distribution reported in Fig. 2d-f. For a $\lambda=$810 nm pump beam polarized at 45 degrees with respect to the crystal  $a$ and $b$ axes, a signature SHG Cherenkov output peaked at $\lambda/2= 405$ nm is detected formed by two TE components in the $a$ direction and two TM components in the $b$ direction, the pump beam being at the center of this diamond-like distribution (Fig. 2d). For a TE pump, two TE components in the $a$ direction are dominant (Fig. 2e), while for a TM pump,  two TM components along the $b$ axis form (Fig. 2f).  Similar results are observed in both samples 1 and 2. An illustration of the SHG experiments for a pump focused into a single polarization-vortex in the $ab$ facet and into an anti-vortex both in the $bc$ and $ac$ facets is reported in Fig. 2g. The situation for a pump focused onto the $ac$ facet is reported in Fig. 2h.  Here, only the $b$ oriented ferroelectric tetrahedrals can contribute to giant refraction Cherenkov SHG, and this only for the TM polarization, a condition that is achieved focusing the pump on an anti-vortex as opposed to a vortex.  The output structure preserves the TM polarization and has a greatly enhanced output angular spectrum that no longer manifests localized peaks.  A similar situation occurs also for light focused onto the $bc$ facet, as reported in Fig. 2i, where the output SHG is emitted at all available angles (see Fig. 2i inset photographs).  The observed SHG follows the basic giant refraction SHG Cherenkov mechanism illustrated  in Fig. 1b (see Cherenkov SHG Experiment Section in Methods).

\begin{figure}[!ht]
\centering
\includegraphics[width=1.0\columnwidth]{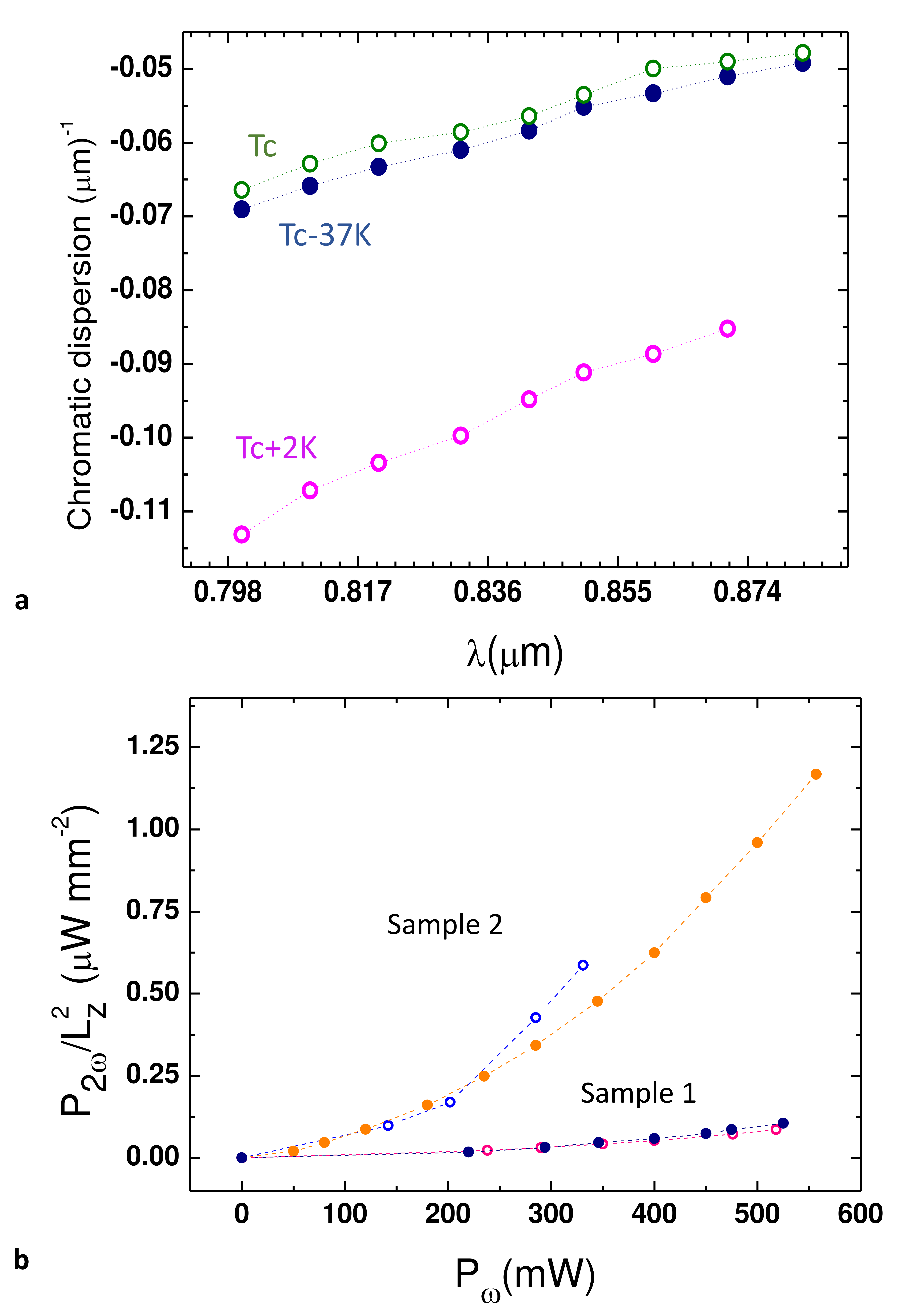}
\caption{\textbf{Chromatic dispersion and output SHG power vs. input pump power.} (a) Supercrystal versus cubic phase chromatic dispersion measured using group velocity dispersion experiments in sample 2. (b) Output SHG power $P_{2\omega}$ normalized to $L_z^2$ versus input pump $P_{\omega}$ at a $\lambda =810$nm and $\theta_0=0$ for sample 1 and sample 2.}
\label{figure3}
\end{figure} 

Results on Cherenkov SHG reported in Fig.2, these including conversion for light propagating along all three principal crystal axes, provide a nonlinear corroboration of evidence of a 3D ferroelectric lattice associated to transmission microscopy \cite{Pierangeli2016, DiMei2018} and polarization transmission microscopy \cite{Ferraro2017}. They also provide, through a direct measurement of $\theta_C$, an estimate of the supercrystal 2$\Delta n n_{2\omega}=(\sin{\theta_C})^2$ (see Cherenkov SHG Experiment Section in Methods).  In the case of Fig.2b, 2$\Delta n n_{2\omega} \simeq 0.08$.  Snell refraction experiments in this direction provide $n_{2\omega}> 26$, so that we expect a $\Delta n < 0.001$, corresponding to an ultra-low approximate dispersion of $dn/d\lambda <-0.002 \mu\text{m}^{-1}$.  The prediction fits in well with our understanding of the supercrystal phase, for which chromatic dispersion is expected to be strongly reduced.  To investigate this further we directly measured supercrystal chromatic dispersion using group-velocity dispersion \cite{Bor1985} for $T>T_C$, where no supercrystal forms, and $T<T_C$, where the supercrystal forms.   Results are reported in Fig.3a.  As expected, the onset of the supercrystal structure is accompanied by a sharp reduction in average values of dispersion, from $dn/d\lambda \simeq -0.10 \mu\text{m}^{-1}$ to $dn/d\lambda \simeq -0.06 \mu\text{m}^{-1}$. This, in turn, is not sufficiently small to circumvent the need for Cherenkov phase-matching ($\theta_{C} \simeq$ 0.28 in Figs. 2d-f).  Strong SHG conversion does not allow a local vortex and anti-vortex dispersion measurement. 

We compared supercrystal SHG in the two samples to identify possible growth and composition related effects.  We found that sample 1 and 2 manifest the same geometrical behavior as regards to giant refraction and Cherenkov SHG, while their net SHG conversion efficiency is considerably different, as reported in Fig.3b. This may be connected to the different values of Curie temperature and/or different values of $\Lambda$ (70 $\mu$m for sample 1 and 50 $\mu$m for sample 2). An estimate of the effective $\chi^{(2)}_{GR}$ is provided in the $\chi^{(2)}_{GR}$ Evaluation Section in Methods. 

\subsection{Spectral and angular acceptance.}
\noindent
%\textbf{Spectral and angular acceptance.} 
The angle at which Cherenkov phase-matching is achieved is wavelength-dependent ($\theta_C(\lambda)$). To characterize this  we report in Fig. 4a measurements of spectral acceptance for a detector able to collect light only from a limited cone at two fixed angles $\theta_1$ (yellow dots) and $\theta_2$ (magenta dots). The result is a spectral bandwidth whose peak follows $\theta_C(\lambda)$ and whose width is in agreement with the angular acceptance (see Angular versus Wavelength Acceptance Section in Methods). Since giant refraction allows no diffraction or pump-signal walk-off, Cherenkov phase-matching will occur for all wavelengths.  In turn, not all Cherenkov SHG can actually leave the output facet of the sample, as total internal reflection occurs for wavevectors that have an internal incidence angle $\theta_i'>1/n_{2\omega}$ with the output facet.  Hence, for a $\theta_i=0$, $\theta_i'=\theta_C'=\arccos{(n_\omega/n_{2\omega})}$,  a zero emitted SHG will result for $\arccos{(n_\omega/n_{2\omega})}>1/n_{2\omega}$.  The effect can be appreciated recalling the full angle-integrated measurement reported in Fig.1h (blue circles).  For a given pump wavelength, the same effect will occur as a function of $\theta_0$ ($\theta_i$): assuming the previously evaluated $\sqrt{2\Delta nn_{2\omega}} \simeq 0.28$, we expect to observe a total internal reflection for an input $|\theta_0|>46^{\circ}$ (see Total Internal Reflection Section in Methods). Measured values of $\theta_C(\lambda)$ are reported in Fig.4b and are in agreement with chromatic dispersion results of Fig.3a. In Fig. 4c we report SHG output, for  a $\lambda=810$nm pump, for different pump launch angles, as in Fig. 1i, but distinguishing between the two Cherenkov components $+\theta_C$ (violet circles) and $-\theta_C$ (magenta circles). SHG suppression is observed for $|\theta_0| >  25^{\circ}$. Illustration of the geometry leading to SHG suppression caused by total internal reflection of the Cherenkov radiation is reported in Fig. 4d.  Once again, the broad spectral and angular acceptance underline how the Cherenkov mechanism in action is not Bragg in nature nor does it relate to quasi-phase-matching.

\begin{figure*}[!ht]
\centering
\includegraphics[width=2\columnwidth]{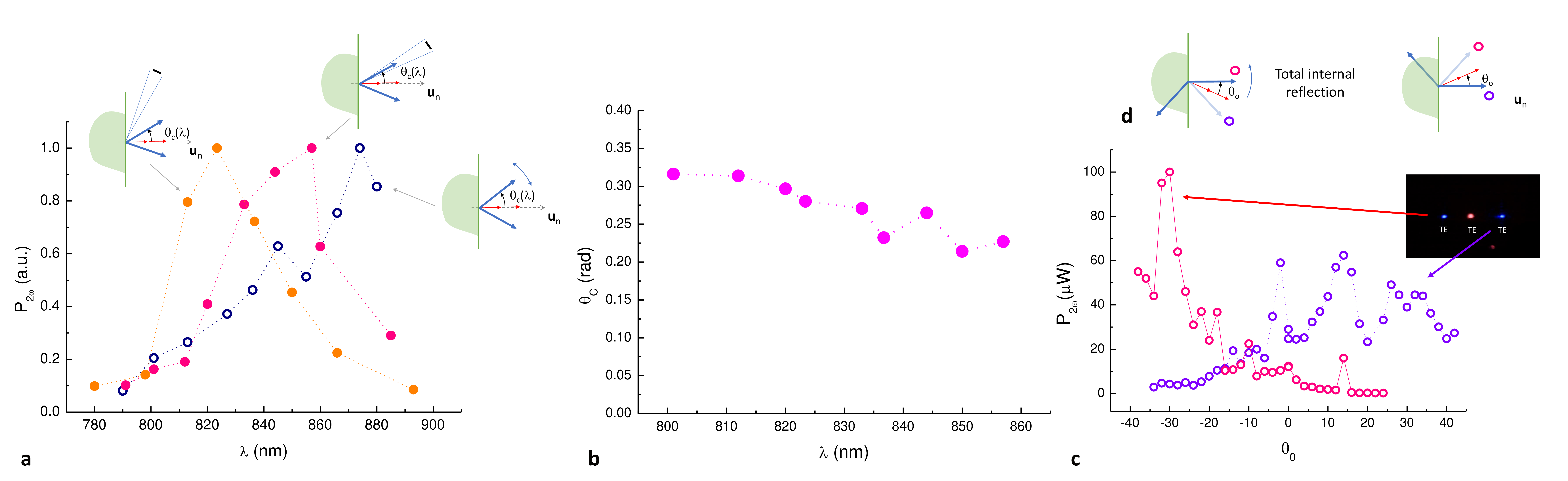}
\caption{\textbf{Cherenkov spectral and angular acceptance.}  (a) Spectral acceptance for a detector placed at two fixed angles (yellow and magenta circles) compared to the super-broad spectral acceptance capturing all emitted light (blue circles). (b) Observed Cherenkov angle versus wavelength. (c) Angular acceptance considering the two TE Cherenkov radiation beams separately (magenta and violet circles). (d) Illustration of the geometry leading to SHG suppression caused by total internal reflection of the Cherenkov radiation. }
\label{figure4}
\end{figure*} 
\subsection{Enhanced Fresnel reflection and extreme nonlinearity}
\noindent
%\textbf{Enhanced Fresnel reflection and extreme nonlinearity.} 
The $n\gg1$ regime leading to SHG (as discussed in Fig. 2) is accompanied by strong Fresnel reflection at the input and output facets. This does not allow a direct evaluation of enhanced wavelength conversion occuring inside the sample by detecting the converted light transmitted outside the sample. Fresnel reflection can be measured directly for the pump, that experiences a conventional $R \simeq$ 0.2, compatible with an average index of refraction $\sim$ 2.6, as expected for light focused onto the vortex and antivortex core \cite{DiMei2018}. By aligning the pump in different positions, a maximum is observed $R \simeq$ 0.45 compatible with an average $n\sim 5$. In these conditions, conversion can be analyzed through the observation of beam dynamics along the propagation direction.  In a standard system, this leads an evolution governed by the so-called Manley-Rowe relationships \cite{Armstrong1962,Boyd2008}. In Fig.5a, b we report scattered SHG light from the body of the sample. Observed scattering in our experiment disappears altogether only when the sample is heated above the Curie point $T_C$ (Fig.5c) and for $T<T_C$ leads to an almost constant SHG signal from the input facet of the sample to the output facet. Analysis of the scattered light versus propagation distance in the sample for different temperatures is reported in Fig. 5d.  The transition from SHG to supercontinuum generation is achieved by doubling the input pump numerical, as reported in Fig. 5e, f (see the Supercontinuum Generation Section in Methods). The origin of this broadband emission and its relation to specific nonlinear processes is still unclear. We recall that SHG with remarkably wide angular and spectral acceptance can be observed in multidomains ferroelectrics \cite{Molina2008, Molina2009}. The enhanced tunability in these studies is associated to phase-matching supported by the underlying disordered domains structure, while in our experiments, phase-mismatch (the transverse component $\Delta k$ in Fig.1) persists. The tunability here is then a product of giant refraction that does not involves Bragg scattering. It serves to further compare our findings to highly versatile SHG in nonlinear photonic crystals \cite{Mateos2012, Li2012}. Here the physical origin of efficient SHG is phase-matching mediated by vector components of the reciprocal lattice in the linear or nonlinear response. It follows that the extent of the pump beam must be larger than the lattice constant. In our case the 15 $\mu$m beam hardly occupies even a single lattice site ($\Lambda \simeq$ 50 $\mu$m).

\section*{Conclusions}
The $n\gg1$ regime forwards a wide range of hereto unobserved and highly versatile nonlinear effects that side other pioneering experiments, such as mismatch-free  nonlinear propagation in zero-index materials \cite{Suchowski2013}. In this paper we have reported our investigation of SHG in conditions of giant refraction. The converted light appears in the form of Cherenkov radiation even in the presence of phase-mismatch. 
This reduces constraints on launch angle, a feature that can considerably mitigate alignment requirements in nonlinear-based light sources. Furthermore, the SHG manifests increased tolerances in wavelength and polarization, a property that can be implemented to  support multiple simultaneous nonlinear processes, with specific impact, for example, in the conversion of infrared images to the visible spectrum.

\begin{figure*}[!ht]
\centering
\includegraphics[width=2\columnwidth]{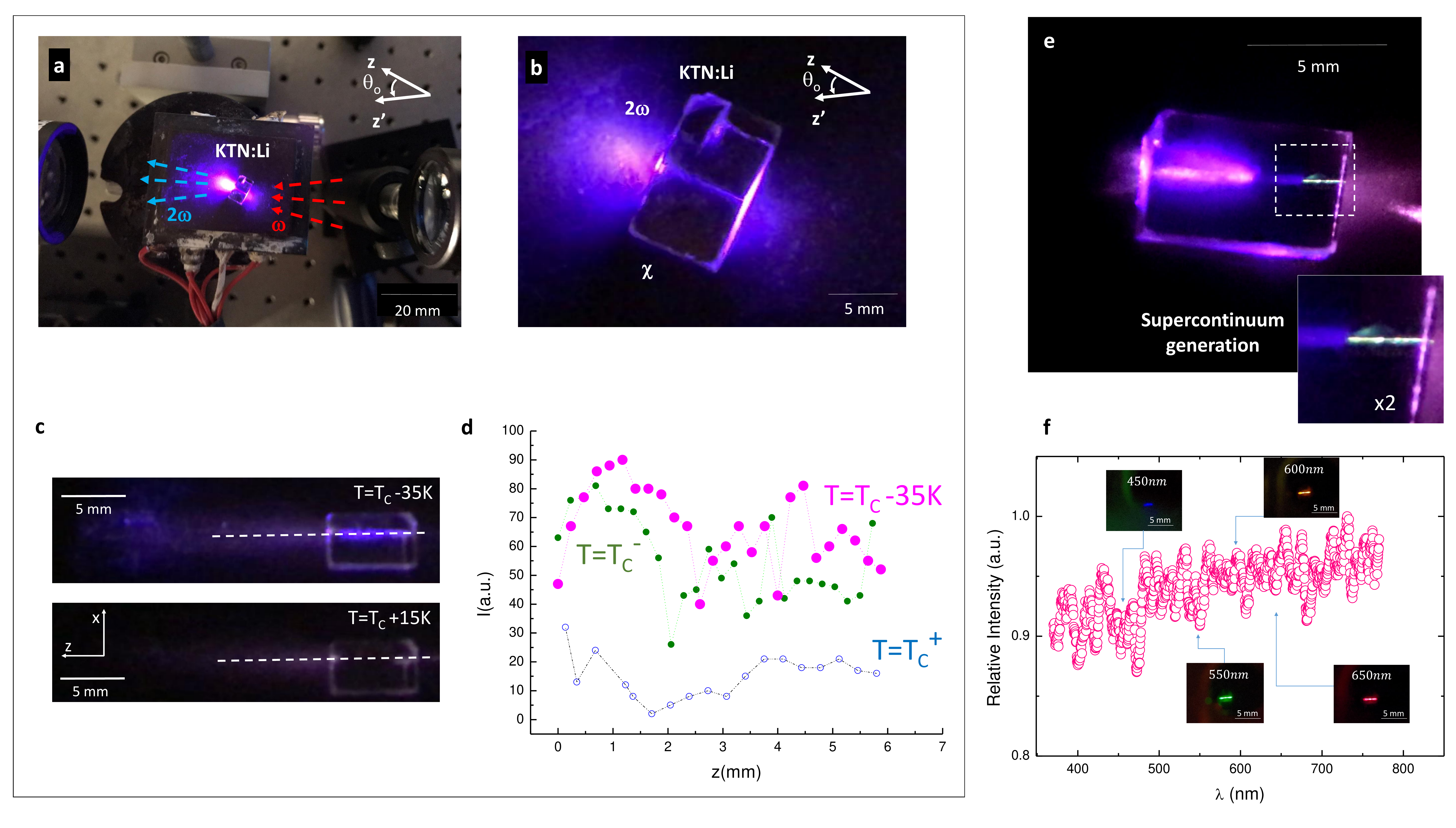}
\caption{\textbf{Strong SHG conversion versus limited net conversion efficiency.}  (a), (b) Top view of pump and SHG signal scattered light ($T=T_C-35K$, sample 2). (c) Scattered light from the body of sample 2 of SHG signal, almost constant for $T<T_C$ from the input facet of the sample to the output facet, while disappears  when the sample is heated above the Curie point $T_C$. (d) Analysis of the scattered light versus propagation distance in the sample for different temperatures indicates a characteristic absence of propagation dynamics for the ferroelectric case ($T=T_C-35K$, magenta full circles), for the Curie temperature on heating from the ferroelectric phase ($T_C^-$, heating from $T_C-35K$ to $T_C$) and on cooling from the paraelectric phase ($T_C^+$, cooling from $T_C+10K$ to $T_C$). (e) Evidence of supercontinuum generation (see Supercontinuum Generation Section in Methods). For further details see Video in Supplementary Movie. (f) Spectrum and multispectral images ($\sim$ 10 nJ/pulse pump at input).}
\label{figure5}
\end{figure*}

\section*{Methods}

\noindent\textbf{Giant refraction Cherenkov phase-matching}. In a material with giant broadband refraction, $n_{\omega},n_{2\omega} \gg 1$.  At the sample input facet the plane-wave components of the pump refract according to the Snell law $\theta_r=\arcsin{(\sin{\theta_i}/n_{\omega})\simeq 0}$,  where $\theta_i$, $\theta_r$ are the incidence and refraction angle.  Cherenkov phase-matching occurs for SHG wavevectors at an angle relative to the pump $\theta_C'=\arccos{(n_{\omega}/n_{2\omega})}\simeq 0$, insomuch that $\Delta n=(n_{2\omega}-n_{\omega})/n_{\omega}\ll 1$, even for a finite $n_{2\omega}-n_{\omega}$.  

\noindent\textbf{Giant refraction Cherenkov SHG}.    SHG polarization is $P_i^{2\omega}=d_{ijk}E_j^{\omega}E_k^\omega$, where $\textbf{E}^\omega$ is the pump field and $d_{ijk}$ is the nonlinear optical susceptibility tensor, that has nonzero components $d_{31},d_{33}$, and $d_{15}$  for the tetragonal 4mm symmetry of KTN:Li \cite{Yariv2002}.  Considering the TE case for a spontaneous polarization parallel to the optical polarization (along the y axis), $P_{\mathit y}^{2\omega}=d_{33}(E_{\mathit y}^{\omega})^2$.  The emitted Cherenkov radiation must then have a $\textbf{k}_{2\omega}$ in a plane orthogonal to $\textbf{P}^{2\omega}$, i.e., in the incidence plane (xz) (central panels in Fig.1b).  Analogously for the TM case, in which, for a spontaneous polarization parallel to the optical polarization (along the x axis), the nonlinear polarization is dominated by the x component $P_{\mathit x}^{2\omega}=d_{33}(E_{\mathit x}^{\omega})^2$, so that Cherenkov SHG occurs in the yz plane (right panels in Fig.1b).  In the TM case, a SHG contribution arises also for a domain with a spontaneous polarization along the z axis, i.e.,  $P_{\mathit x}^{2\omega}=2d_{15}E_{\mathit x}^{\omega}E_{\mathit z}^{\omega}$ and $P_{\mathit z}^{2\omega}=d_{31}(E_{\mathit x}^{\omega})^2+d_{33}(E_{\mathit z}^{\omega})^2$.  The emitted Cherenkov SHG will then be TM polarized and have a $\textbf{k}_{2\omega}$ orthogonal to $\textbf{P}^{2\omega}$ in the incidence plane xz (i.e., along the y axis). This situation is particularly relevant for results reported in Fig.2h and Fig.2i. In this case the spontaneous polarization is oriented orthogonal to the input facet, so that while the pump is prevalently experiencing a standard index of refraction $n_\omega$, the SHG is dominated by $d_{33}$ and has a stronger component along the direction of spontaneous polarization. The result then is that $n_\omega/n_{2\omega}\ll 1$, so that $\theta'_{C}$ inside the sample remains finite, while all waves still have their Poynting vectors along the normal to the input facet (giant refraction). This causes this angular amplification caused by the angular spectrum of the focused pump to populate in a continuous manner a wide angle of SHG emission around the pump average propagation direction that, on output, can ever occupy the entire angular spectrum ($2 \theta_C =\pi$, Fig. 2i).

\noindent\textbf{Materials}.
The two samples (sample 1 and sample 2) are zero-cut polished lithium-enriched solid-solutions of potassium-tantalate-niobate (KTN:Li). They have the same composition K$_{0.997}$Ta$_{0.64}$Nb$_{0.36}$O$_3$:Li$_{0.003}$, while in the flux of sample 2 traces of Mo impurities are introduced. Sample 1 measures along its three axes 4.62$^{\mathit (a)}$ x 3.86$^{\mathit (b)}$ x 1.6$^{\mathit (c)}$ mm while sample 2 is 6.96$^{\mathit (a)}$ x 3.86$^{\mathit (b)}$ x 1.6$^{\mathit (c)}$ mm.  The samples form perovskites with room-temperature cubic-to-tetragonal (m3m to 4mm) ferroelectric phase-transition temperatures $T_{C,1}=315$ K and $T_{C,2}=333$ K. Both are grown through the top-seeded method that causes them to have a built-in spatially periodic oscillation in composition along the growth axis (the $a$ axis) that translates into an approximately periodic $\Lambda=50$ $\mu$m striation grating (for sample 2) that then determines the lattice constant of the underlying super-crystal \cite{Pierangeli2016}.

\noindent\textbf{Supercrystal Preparation}.
Each sample, initially equilibrated at $T=298$K and unbiased, is heated to $\simeq 373$ K at a rate of 0.6 K/min and is DC-biased by an electric field that increases at a constant rate from 0 to 4 kV cm$^{-1}$.  The sample is then cooled back down to $T=298$K while the bias field remains constant at 4 kV cm$^{-1}$. The DC field is applied between the two parallel faces along the $a$ axis (growth axis). To minimize the temperature gradient, the sample is dipped into a Teflon holder that contains temperature resistant mineral oil. The supercrystal can now be further modified having the sample undergo successive thermal cycles, composed of a first stage in which the unbiased sample is heated to $T_C+10 K$ at a rate of $0.35$K$s^{-1}$ immediately followed by a second cooling stage tp $T_C-35$K at a rate of $0.1$ K$s^{-1}$.  Once the thermal protocol in completed, each sample is used for optical experiments at a given temperature $T<T_C$.

\noindent\textbf{Giant refraction Experiments}. The sample is cooled using a current-controlled Peltier junction to $T_C-35$K and rotated by a tunable angle $\theta_0$ with respect to the optical propagation axis $z'$ (see scheme illustrated in Fig.1c).  Light is collected from a commercial projector (NEC-VE281X, XGA, 2800 lumens) polarized using a linear polarization filter and focused onto the input facet of the sample using a  high-aperture long-working distance microscope objective (Edmund Optics, x100, 3mm
working distance, achromatic, NA$=0.8$) positioned $\simeq$ 30 cm from the output lens of the projector.  The top-view image in Fig. 1d is taken using an  Apple iPhone7. Top-view scattered light from within the sample and from the lower metallic support indicates strong reflection from the input facet and a non-spreading propagation inside the sample normal to the input facet irrespective of wavelength, $\theta_0$, and launch polarization, and a regular diffraction of the beam exiting the sample, as expected for giant refraction.  %Crystal positions that lead to GR in the samples are displaced along the transverse $x$ and $y$ directions by 54-60 $\mu$m depending on the region inspected.

\noindent\textbf{SHG Setup}.
SHG experiments (see scheme and photo of apparatus in Fig. 1e) are carried out in the 790-880 nm range using a Tsunami Spectra Physics Ti:Sa CW mode-locked laser (maximum output power of 0.6W at $\lambda = 810 \pm 7$ nm), with a repetition rate of 80 MHz and a pulsewidth of 190 fs.  Laser beam linear polarization, TM or TE, or a superposition of the two, is set using a $\lambda /2$ waveplate. The beam is focused onto the input facet of the $\theta_0$-rotated sample using a 50-mm-focal-length lens. The pump beam is focused to an input FWHM  $\simeq 15 \mu$m. The SHG pattern is detected on a white screen placed at $d$= 7.0 cm from the output facet of the sample using a Canon EOS 50d. SHG power $P_{2\omega}$ is measured in Fig. 1g (and Fig. 3b) filtering and focusing converted light onto a power meter for a TM pump (and SHG). 

\noindent\textbf{Acceptance}.  Spectral acceptance is reported for $\theta_0=0$ in arbitrary power units $P_{2\omega}$ normalized to the peak spectral value.  Since each measurement at different wavelengths is carried out with different pump power, the output signal is rescaled appropriately, i.e., divided by the input power squared. Angular acceptance is evaluated for a $810$nm pump and for all accessible launch angles.  In both spectral and angular acceptance experiments, output SHG is collected by a lens and focused onto a power meter. Note that on consequence of giant refraction, the effective propagation length in the sample is launch-angle independent and equal to the length of the sample in the propagation direction (see Fig. 1b). Wavelength dependence, due to the Fresnel reflection at output, is analyzed in Fig.4. 
% As discussed in relation to the phase-matching mechanism, SHG depends strongly on the position of the sample in the xy plane relative to the input pump beam. Experiments reported in Fig.1D and Fig.1E include results in which this position is optimized for conversion, while in Fig.1F the sample is solely rotated, the result being strong oscillations in observed $P_{2\omega}$. 

\noindent
\textbf{Cherenkov SHG Experiment.} 
$\cos{(\theta_C')}=(2k_{\omega}/k_{2\omega})=n_{\omega}/n_{2\omega}$. For normal dispersion ($n_{2\omega}>n_{\omega}$) $n_{\omega}=n_{2\omega}-\Delta n$, so that since $\theta_C'\ll 1$,  $\theta_C'\simeq \sqrt{2\Delta n/n_{2\omega}}$.     Outside the sample, $\sin{(\theta_C)}=n_{2\omega}\sin{\theta_C'}\simeq n_{2\omega}\theta_C'$.  Measuring $\theta_C$ leads to an estimate of $\Delta n=(\sin{(\theta_C)})^2/(2n_{2\omega})$.  For a pump focused on the $ab$ facet in Fig.2b, for both the TE and TM cases, the two SH beams emerge in the x-z (i.e., $ac$) and y-z (i.e., $bc$) planes at an angle $\theta_C \simeq 0.28$ rad with respect to the pump (for all accessible values of $\theta_i$).  According to the Cherenkov model, this implies that $\sqrt{2\Delta n n_{2\omega}}\simeq 0.28$.  For light focused on the $ac$ facet in Fig. 2h and $bc$ facet in Fig. 2i, SHG is generated from tetrahedral domains oriented along the $b$ and $a$ axis, respectively, i.e., with a spontaneous polarization orthogonal to the input facet.  Involving both $d_{31},d_{33},d_{15}$, the result is a TM SHG in the $bc$ and $ac$ plane, respectively.  The observed $\theta_C \simeq \pi/2$ (see detailed photos in the inset of Fig. 2i), corresponding to a $\sqrt{2\Delta n n_{2\omega}}\simeq 1$. A similar contribution associated to $d_{15}$ in the case reported in Fig. 2b, c, where $d_{33}$ contributions are dominant, and leads to the ``spurious" SHG scattering in the TM case (Fig. 2f) as opposed to the TE case (Fig. 2e).

\noindent\textbf{\bm{$\chi^{(2)}_{GR}$} Evaluation}.  
To provide an estimate of the effective nonlinear susceptibility $\chi^{(2)}_{GR}$ we can make use of the simplified plane-wave model (diffraction is absent in the GR regime) described in Ref. \cite{Boyd2008}, pages 77-79. For sample 1, the time averaged powers at the output are $P_{\omega}= 510 mW$ and $P_{2\omega}= 2 \mu W$ for the 190 fs pulse train operating at 80 MHz repetition rate, while the beam waste is taken to be $w_{0} \simeq 15 \mu m$. For the purpose of the evaluation of peak intensity and Fresnel reflection, the regions leading to SHG for the TM case have $n_{\omega} \sim n_{2\omega} \sim n_{GR}$, where 
$n_{GR}\gg$1. Of the input pump beam, only a portion actually interact with the tetrahedral structures allowing SHG (the $d_{33}$ solids in Fig.2b). The fraction of active area can be evaluated measuring the Fresnel reflection at input and output facets, comparing it to what expected for regions with standard reflection (i.e., for $n \simeq 2.2$) and those regions that have $d_{33}$ and hence an enhanced reflection associated to $n_{GR}$. While longitudinal phase-matching is guaranteed by the Cherenkov-like geometry (Fig.1a), the aperture of the pump and SHG beams remains finite ($\Delta \theta '\sim\Delta\theta_{ext}/n_{GR}$, where $\Delta \theta_{ext} \simeq$ 0.07). This introduces a residual longitudinal mismatch. The transverse wavevector mismatch can be evaluated considering $\theta_{c}\simeq$ 0.28 rad measured outside the sample, leading to $\Delta k \simeq 4.4 {\mu m}^{-1}$. The result is $\chi^{(2)}_{GR} \simeq 5.2$ pm V$^{-1}$ ${n^{2}}_{GR} $. Considering even the minimum value of $n_{GR}$ as measured from diffraction and Snell refraction ($n_{GR} \simeq$ 26), we obtain an effective  $\chi^{(2)}_{GR} \simeq 3.5 \times 10^3$ pm V$^{-1}$, to be compared to the measured value of standard KTN, i.e., $\chi^{(2)}_{KTN} \simeq 168$ pm V$^{-1}$ (n$\simeq 2.3$) \cite{Zhang1997}.

\noindent\textbf{Angular versus Wavelength Acceptance}.  To test this we maximized SHG efficiency, i.e., Cherenkov phase-matching is established for the specific pump wavelength $\lambda$, and the SHG signal detector is placed so as to capture a single output diffraction-limited mode.  As reported in Fig.4a, changing the pump wavelength without altering the crystal and detector geometry leads to a relative spectral acceptance $\Delta \lambda/ \lambda \simeq 0.047$  that is in agreement with the input pump numerical aperture $2\lambda/ \pi w_{0} \simeq 0.05$.

\noindent\textbf{Total internal reflection}.  Total internal reflection of the SHG signal occurs at the output facet when approximately $\theta_C'+\theta_r>1/n$, where $\theta_r=\sin{(\theta_0)}/n$.  Assuming that $\theta_C'=\sqrt{2\Delta n/n_{2\omega}}$, we have that total internal reflection occurs for $|\sin{(\theta_0)}|>1-\sqrt{2\Delta n n_{2\omega}}$.  Taking the value of $\sqrt{2\Delta nn_{2\omega}}\simeq 0.28$ gives $|\theta_0|>46^{\circ}$.

\noindent\textbf{Supercontinuum generation}. Experiments are carried out replacing the 50mm lens with a 25mm one. The pump is now focused in proximity of the input facet of sample 2 (Fig. 5e) and a characteristic white plume is detected.

\section*{Data availability }
\noindent
The data that support the plots within this paper and other findings of this study
are available from the corresponding author upon reasonable request.

\noindent\textbf{Acknowledgements}. This research was supported in part by the Israel Science Foundation (Grant No. 1960/16). F.D. and E.D. were supported through the \textsc{Attract} project funded by the EC under Grant Agreement 777222 and the Sapienza Ricerca di Ateneo 2019 project. \\%
\noindent\textbf{Supplementary Information}. Supplementary Information is linked to the online version of the paper at www.nature.com/nature.\\%
\noindent\textbf{Author Contribution}. A.J.A. developed and synthesized the materials. L.F., L.T., F.D., M.F., J.P., D.P. carried out SHG measurements. L.F. and L.T. carried out chromatic dispersion measurements. GB.P. carried out the SC poling procedure.  E.D., F.D., L.F., and P.D. elaborated the physical framework and interpretation. L.F., L.T., F.D., M.F., J.P., D.P., GB.P., F.X., P.D., A.J.A., and E.D. participated in discussions.  E.D., L.F., and F.D. wrote the article with the help of all authors.\\%
\noindent\textbf{Competing Interests}. The authors declare that they have no competing financial and non-financial interests. \\%
\noindent\textbf{Reprints}. Reprints and permissions information is available at npg.nature.com/reprintsandpermissions.\\%
\noindent\textbf{Correspondence}. *Correspondence and requests for materials should be addressed to L.F.~(email:   ludovica.falsi@uniroma1.it)\\


\vspace*{0.2cm}



\begin{thebibliography}{25}
 

\bibitem{Boyd2008} %12
Boyd, R.W., \textit{Nonlinear Optics}, 3rd ed.,  (Academic Press, 2008).

\bibitem{Shen1984} Shen, Y. R., \textit{The Principles of Nonlinear Optics} (Wiley-Interscience, New York, 1984). 

\bibitem{Loudon2010} %12
Loudon, R., \textit{The Quantum Theory of Light}, 3rd ed.,  (Oxford Science Publications, 2010).

\bibitem{Brown2003} Brown, E., McKee, T.,  diTomaso, E., Pluen, A.,  Seed, B.,  Boucher, Y., \& Jain, R. K., ``Dynamic imaging of collagen and its modulation in tumors in vivo using second-harmonic generation,'' \emph{Nat. Med.} \textbf{9}, 796-800 (2003). 

\bibitem{Tiecke2014} Tiecke, T. G., Thompson,  J. D., de Leon, N. P., Liu, L. R., Vuletic, V., \& Lukin, M. D. ``Nanophotonic quantum phase switch with a single atom,'' \emph{Nature} \textbf{508}, 241 (2014). 



\bibitem{Reiserer2014} Reiserer, A., Kalb, N.,  Rempe, G., \&  Ritter, S. ``A quantum gate between a flying optical photon and a single trapped atom,'' \emph{Nature} \textbf{508}, 237 (2014). 

\bibitem{Chang2014} Chang, D. E., Vuletić, V.  \& Lukin, M. D. ``Quantum nonlinear optics - photon by photon,'' \emph{Nat. Photon.}  \textbf{8}, 685-694 (2014).

\bibitem{Wei2018} Wei, D., Wang, C., Wang, H.,  Hu, X., Wei, D., Fang, X.,  Zhang, Y., Wu, D.,  Hu, Y.,  Li, J.,  Zhu, S., \& Xiao, M., ``Experimental demonstration of a three-dimensional lithium niobate nonlinear photonic crystal,'' \emph{Nat. Photon.} 12, 596-600 (2018). 

\bibitem{Xu2018} Xu, T.,  Switkowski, K., Chen, X., Liu, S., Koynov, K., Yu, H.,  Zhang, H.,  Wang, J.,  Sheng, Y., \& Krolikowski, W., ``Three-dimensional nonlinear photonic crystal in ferroelectric barium calcium titanate,'' \emph{ Nat. Photon.} 12, 591-595 (2018).

\bibitem{Zhang2019} Zhang, X., Yang, Q.X., Liu, H.L., Wang, X.P., He, S., Li, X.J., Wu, P.F.
``Switching effects of spontaneously formed superlattices in relaxor ferroelectrics,'' \emph{Opt. Mat. Express} \textbf{9}, 4081-4089 (2019).

\bibitem{Stoica2019} Stoica, V. A. , Laanait, N. , Dai, C., Hong, Z., Yuan, Y., Zhang, Z., Lei, S., McCarter, M. R., Yadav, A.,  Damodaran, A. R., Das, S., Stone, G. A., Karapetrova, J., Walko, D. A., Zhang, X. , Martin, L. W., Ramesh, R., Chen, L.-Q., Wen, H., Gopalan, V., \& Freeland, J. W.  ``Optical creation of a supercrystal with three-dimensional nanoscale periodicity,'' \emph{Nat. Mater.} \textbf{18}, 377-383 (2019).

\bibitem{Liu2019}  Liu, S.,  Switkowski, K., Xu, C., Tian, J., Wang, B., Lu, P., Krolikowski, W., \&  Sheng, Y., ``Nonlinear wavefront shaping with optically induced three-dimensional nonlinear photonic crystals,'' \emph{Nat. Commun.} 10, 3208 (2019).


\bibitem{Jelley1958} %12
Jelley, J.V., \textit{Cerenkov Radiation}, (Pergmon Press, 1958).


\bibitem{Mathieu1969} Mathieu, E. ``Conditions for quasi Cerenkov radiation, generated by optical second harmonic polarisation in a
nonlinear cristal,'' \emph{Journal of Applied Mathematics and Physics}(ZAMP) \textbf{20}, 433–439 (1969).

\bibitem{Tien1970} 
Tien, P. K., Ulrich , R.,\& Martin, J.  ``Optical second harmonic generation in form of coherent Cerenkov radiation from a thin-film waveguide,'' \emph{Appl. Phys. Lett.} \textbf{17}, 447 (1970).

\bibitem{Zhang2008} Zhang, Y., Gao, Z. D., Qi, Z., Zhu, S. N. , \& Ming, N. B. ``Nonlinear Čerenkov Radiation in Nonlinear Photonic Crystal Waveguides,'' \emph{Phys. Rev. Lett. } \textbf{100}, 163904 (2008).

\bibitem{Sheng2010} 
Sheng, Y., Best, A., Butt, H.-J., Krolikowski, W., Arie, A.,    \& Koynov, K. ``Three-dimensional ferroelectric domain visualization by Cerenkov-type second harmonic generation,'' \emph{Opt. Exp.} \textbf{18}, 16539 (2010).

\bibitem{Sheng2012} 
Sheng, Y., Kong , Q. , Roppo, V. , Kalinowski , K.,  Wang, Q., Cojocaru, C., \& Krolikowski, W.  ``Theoretical study of Čerenkov-type second-harmonic generation in periodically poled ferroelectric crystals,'' \emph{J. Opt. Soc. Am. B} \textbf{29}, 312-318 (2012).


\bibitem{Roppo2013} 
Roppo, V., Kalinowski , K., Sheng, Y., Krolikowski, W.,  Cojocaru, C., \& Trull, J.  ``Unified approach to Cerenkov second harmonic generation,'' \emph{Opt. Express} \textbf{21}, 25715-25726 (2013).



\bibitem{Ni2016} Ni, R., Du, L., Wu, Y., Hu, X. P., Zou, J., Sheng, Y., Arie, A., Zhang, Y., Zhu, S. N., ``Nonlinear Cherenkov difference-frequency generation exploiting birefringence of KTP," \emph{Appl. Phys. Lett.} 108, 031104 (2016).

\bibitem{Pierangeli2016} Pierangeli, D., Ferraro, M. , Di Mei, F. , Di Domenico, G., De Oliveira, C.E.M., Agranat, A.J. , \& DelRe, E.  ``Super-crystals in composite ferroelectrics,'' \emph{Nat. Commun.} \textbf{7}, 10674 (2016).

\bibitem{DiMei2018} Di Mei, F., Falsi, L.,  Flammini, M.,  Pierangeli, D.,  Di Porto, P.,  Agranat, A. J., \& DelRe, E.  ``Giant broadband refraction in the visible in a ferroelectric perovskite,'' \emph{Nat. Photon.}  12, 734-738 (2018).

\bibitem{Armstrong1962}
Armstrong, J. A., Bloembergen, N., Ducuing, J., \& Pershan, P. S.,
 ``Interactions between Light Waves in a Nonlinear Dielectric,''
Phys. Rev. \textbf{127}, 1918 (1962).


\bibitem{Kleinman1962}
Kleinman, D. A.,
 ``Theory of Second Harmonic Generation of Light,''
Phys. Rev. \textbf{128}, 1761 (1962).

\bibitem{Fejer1992} Fejer, M. M., Magel, G. A., Jundt, D. H., \& Byer, R. L., ``Quasi-phase-matched second harmonic generation: tuning and tolerances," \emph{J. Quant. Electron.}  \textbf{28}, 2631-2654 (1992).

\bibitem{Saltiel2009} Saltiel, S. M. , Neshev, D. N., Krolikowski, W.,  Arie, A.,  Bang, O., \&  Kivshar, Y. S., ``Multiorder nonlinear diffraction in frequency doubling processes," \emph{Opt. Lett.} \textbf{34}, 848-850 (2009).

\bibitem{Wang2017} Wang, J, Jin, K., Gu, J.,  Wan, Q.,  Yao, H.,  \&  Yang, G., ``Direct evidence of correlation between the second harmonic generation anisotropy patterns and the polarization orientation of perovskite ferroelectric,'' \emph{Sci. Rep.}  \textbf{7}, 9051 (2017).  

\bibitem{Ferraro2017} Ferraro, M., Pierangeli, D., Flammini,  M., Di Domenico, G., Falsi, L., Di Mei, F., Agranat, A. J., \& DelRe, E. ``Observation of polarization-maintaining light propagation in depoled compositionally disordered ferroelectrics,'' \emph{Opt. Lett.} \textbf{42}, 3856-3859 (2017).

\bibitem{Bor1985} Bor,  Z., \& Rácz, B., ``Dispersion of optical materials used for picosecond spectroscopy," \emph{Appl. Opt.} \textbf{24}, 3440-3441 (1985).




\bibitem{Roppo2010} Roppo, V., Wang, W. , Kalinowski, K. , Kong, Y.,  Cojocaru, C., Trull, J.,  Vilaseca, R.,  Scalora, M., Krolikowski,  W. \&  Kivshar, Y., ``The role of ferroelectric domain structure in second harmonic generation in random quadratic media," \emph{Opt. Express} \textbf{18}, 4012-4022 (2010).


\bibitem{Ayoub2011} Ayoub, M.,  Imbrock, J., \& Denz, C.,  ``Second harmonic generation in multi-domain $\chi^2$ media: from disorder to order," \emph{Opt. Express} \textbf{19}, 11340-11354 (2011).

\bibitem{Molina2008} Molina, P.,  Ramírez, M.d.l.O., \& Bausá, L. E.,  ``Strontium Barium Niobate as a Multifunctional Two‐Dimensional Nonlinear Photonic Glass," \emph{Adv. Funct. Mater.} \textbf{18}, 709-715 (2008).

\bibitem{Molina2009} Molina, P., Álvarez-García, S., Ramírez, M.O., García-Solé, J., Bausá, L. E., et al.  ``Nonlinear prism based on the natural ferroelectric domain structure in
calcium barium niobate," \emph{Appl. Phys. Lett.} \textbf{94}, 071111 (2009).
 
\bibitem{Mateos2012} Mateos, L., Molina, P., Galisteo, J., López, C., Bausá, L. E., \& Ramírez, M. O.,  ``Simultaneous generation of second to fifth harmonic conical beams in a two dimensional nonlinear photonic crystal," \emph{Opt. Express} \textbf{20}, 29940-29948 (2012).
  
\bibitem{Li2012} Li, H. X., Mu,  S. Y., Xu, P., Zhong, M. L., Chen, C. D., Hu, X. P., Cui, W. N., \& Zhu, S. N., ``Multicolor Čerenkov conical beams
generation by cascaded-$\chi^{(2)}$ processes in
radially poled nonlinear photonic crystals," \emph{Appl. Phys. Lett} \textbf{100}, 101101 (2012).
 
\bibitem{Suchowski2013} Suchowski, H., O’Brien, K., Wong, Z. J., Salandrino, A., Yin, X., \& Zhang, X.  ``Phase Mismatch–Free Nonlinear Propagation in Optical Zero-Index Materials,'' \emph{Science}  \textbf{342}, 1223-1226 (2013).



\bibitem{Yariv2002} Yariv, A., \&  Yeh P.,  \textit{Optical Waves in Crystals: Propagation and Control of Laser Radiation},  (Wiley, 2002).

\bibitem{Zhang1997} Zhang, H. Y., He, X. H., Shih, Y. H., Harshavardhan, K. S., \& Knauss, L. A.  ``Optical and nonlinear optical study of KTa$_{0.52}$Nb$_{0.48}$O$_3$ epitaxial film,'' \emph{Opt. Lett.}  \textbf{22}, 1745-1747 (1997).




\end{thebibliography}
\end{document}